\documentclass[aps,prb,reprint,floatfix,amsmath,amssymb, superscriptaddress]{revtex4-1}

\usepackage{graphicx}
\usepackage{epstopdf}
\usepackage{color}
\usepackage{hyperref}
\usepackage{bm}
\usepackage{printlen}
\usepackage{units}

\renewcommand{\vec}[1]{\mathbf{#1}}

\newcommand{\Trace}{\text{Tr}\,}

\hypersetup{hidelinks}

\begin{document}

\title{Re-entrant magic-angle phenomena in twisted bilayer graphene 
\\
in  integer magnetic fluxes}

\author{Yifei Guan}  
\affiliation{Institute of Physics, Ecole Polytechnique Fédérale de Lausanne (EPFL), Lausanne, CH-1015, Switzerland}    

\author{Oleg V. Yazyev}  
\affiliation{Institute of Physics, Ecole Polytechnique Fédérale de Lausanne (EPFL), Lausanne, CH-1015, Switzerland}  

\author{Alexander  Kruchkov}   
\affiliation{Institute of Physics, Ecole Polytechnique Fédérale de Lausanne (EPFL), Lausanne, CH-1015, Switzerland}   
\affiliation{Department of Physics, Harvard University, Cambridge, MA 02138, USA}    
\affiliation{Branco Weiss Society in Science, ETH Zurich, Zurich, Switzerland}

\begin{abstract}
In this work we address the re-entrance of magic-angle phenomena (band flatness and quantum-geometric transport) in twisted bilayer graphene (TBG) subjected to strong magnetic fluxes $\pm \Phi_0$, $\pm 2 \Phi_0$, $\pm 3 \Phi_0$... ($\Phi_0 = h/e$ is the flux quantum per moiré cell). The moiré translation invariance is restored at the integer fluxes, for which we calculate the TBG band structure using accurate atomistic models with lattice relaxations. Similarly to the zero-flux physics outside  the magic angle condition, the reported effect breaks down rapidly with the twist. We conclude that the magic-angle physics re-emerges in high magnetic fields, witnessed by the appearance of flat electronic bands \textit{distinct} from Landau levels, and manifesting non-trivial quantum geometry. We further discuss the possible flat-band quantum geometric contribution to the superfluid weight  in strong magnetic fields (28 T at 1.08$^\circ$ twist), according to Peotta-T\"{o}rm\"{a} mechanism. 
\end{abstract}

\maketitle

In 2D systems, the electronic spectrum in magnetic field develops a fractal structure ("Hofstadter butterfly"),\cite{Hofstadter1976} which lowers the effective dimensionality, and contributes to suppressing superconductivity (long-range order is generically destroyed in dimensions lower than $D$=$2$). The observation of Hofstadter physics requires strong magnetic fluxes ($\sim$$h/e$), which became experimentally accessible only with the advent of moiré superlattices.\cite{Dean2013,Hunt2013,Ponomarenko2013} In those experiments,\cite{Dean2013,Hunt2013,Ponomarenko2013} the magnetic fields of nearly 30 T were employed in the system of graphene monolayer twisted on hexagonal boron nitride (hBN), resulting into effective fluxes of $\Phi = B/A \sim \Phi_0$ ($A$ is the moiré cell area, $\Phi_0$$=$$h/e = $$4$ Wb is magnetic flux quantum). Furthermore, the twisted graphene multilayers provide a natural platform to test the interplay between the Hofstadter physics and strong correlations.\cite{Cao2018, Hao2021, Park2021,Park2021b,Zhang2021}
In twisted bilayer graphene, the smaller is the twist $\theta$, the larger is the effective magnetic flux ($\Phi\propto 1/\theta^2$) at the fixed field $\vec{B}$: for TBG at the magic angle $1.08^{\circ}$ the magnetic flux quantum corresponds to $B_0 \approx 2 8 $ T, which is reachable in the modern laboratories.\cite{Hahn2019}
 
The magic-angle graphene heterostructures---two or more graphene sheets twisted to the angle  $\sim$$1^{\circ}$, at which a very narrow band emerges in the electronic spectrum---have re-attracted significant attention due to re-entrant superconductivity in strong magnetic fields and reported Pauli limit violation.\cite{Cao2021,Chaudhary2021,Shaffer2021} The re-entrant correlated (Chern) insulator phases were reported at strong magnetic fluxes,\cite{Das2021,Arbeitman2021} close to the unit magnetic flux quantum $\Phi_0 = h/e$ per moiré unit cell (see also Ref.\cite{Sheffer2021}). In this paper we show that in all the integer magnetic flux $\Phi = \pm N \Phi_0$, the magic angle physics of TBG is re-entrant and similar to the physics of TBG in zero magnetic field.  Namely, we find non-trivial flat Chern bands  
at integer flux $h/e$ (see Fig. 1). These bands are \textit{distinct} from Landau levels for three reasons: \textbf{(i)} Outside 
the magic angle range 
the band flatness breaks down;  \textbf{(ii)} The magic-angle flat bands are characterized by the total Chern number $|C|=2$, while Landau levels are characterized by $|C|=1$; and \textbf{(iii)} The quantum geometry (Fubini-Study metrics $\mathfrak G_{ij}$)  of the magic-angle flat bands is highly  nontrivial and different from the Landau level case, and provides  the enhanced basis for quantum-geometric transport\cite{Peotta2015} in flat bands without dispersive contributions due to  $\sum_{\vec k} \Trace \mathfrak G_{ij} (\vec k) \approx 3$.

\begin{figure}[b]
\includegraphics[width=1.0 \columnwidth]{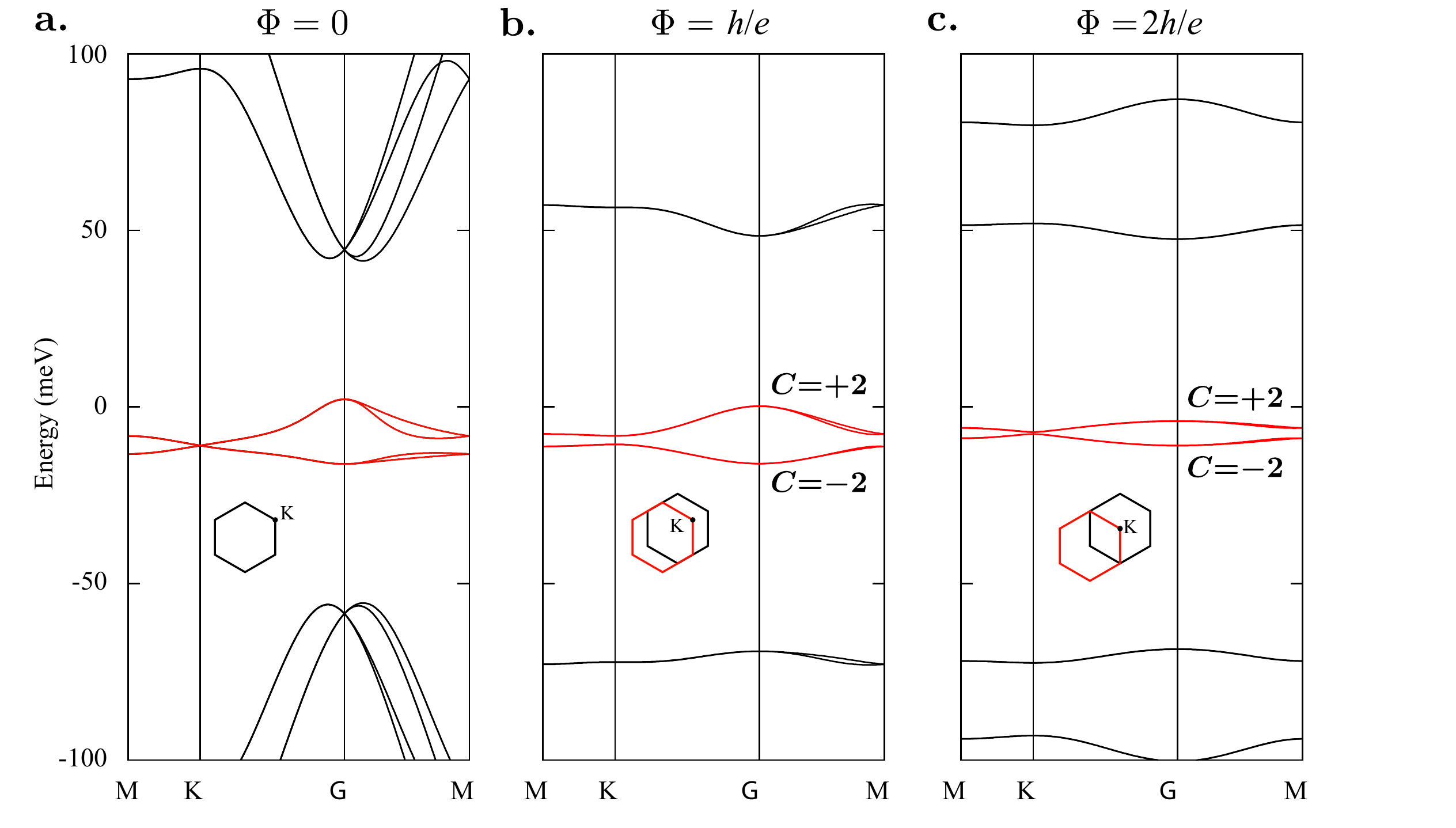}
\caption{Electronic band structure of magic-angle  twisted bilayer graphene in integer magnetic flux:   (a) zero flux ($\Phi = 0$); (b) flux one ($\Phi= h/e$); (c) flux two ($\Phi= 2 h/e$). All band structures are  calculated  with tight binding model  including lattice relaxations effects. In integer magnetic flux, the lower (two out four) flat bands acquire $C=-2$.  }
\end{figure}

In this paper we report that the nontrivial quantum geometry, electronic  band flatness, and conditions for unconventional quantum transport are re-established at integer magnetic flux (in units of $h/e$) through the moiré unit cell. Similarly to the zero-flux case, the Fermi velocity and the bandwidth drop dramatically at the magic angle, to reappear as a dispersive band both below and above the magic angle. Importantly, the flat bands in finite flux has non-trivial Chern numbers $|C|$=$2$ {(defined at half filling, see Fig. 1)} and non-trivial quantum geometry, which follows  the quantum-geometric flatness criterion\cite{Kruchkov2021a}
\begin{align}
\text{Tr} \mathcal G_{ij} (\vec k)  \simeq  \mathcal F_{xy} (\vec k) . 
\label{ideal}
\end{align}
We show numerically with atomistic calculations that in the realistic TBG,  the "ideal band condition" \eqref{ideal} is satisfied  in the magnetic moiré Brillouin zone (mmBZ) regions   where the band flatness is pronounced in terms of vanishing Fermi velocity (e.g. around the K points of mmBZ). Here $\mathcal G_{ij}$ and $F_{xy} (\vec k)$ are the real and imaginary parts of the quantum-geometric tensor $\mathfrak{G}_{ij}$(see further), determining the quantum distance between electronic states in the (projected) Hilbert space.\cite{Provost1980}

In this paper, we investigate the effect of strong magnetic fields with integer flux in twisted bilayer graphene with the help of the accurate atomistic model including lattice relaxation effects at the magic angle, and compare the observed results with the established knowledge of the zero-flux TBG case. The details of the tight-binding Hamiltonian are provided in Supplementary Materials (SM).\cite{SI} The key observation is that the magnetic translation operators commute at every integer flux $\Phi = N \Phi_0$, namely
\begin{align}
\hat T_{\vec a_1} \hat T_{\vec a_2}  = e^{i 2 \pi \Phi/\Phi_0} \hat T_{\vec a_2} \hat T_{\vec a_1}, \ \ \to \ \ [\hat T_{\vec a_1}, \hat T_{\vec a_2}]_{\Phi = N\Phi_0}  = 0. 
\nonumber
\end{align}
  Thus the moiré unit cell  is restored, and the system flows towards the electronic band structure defined on the mmBZ, which in the integer flux has the same periodicity as the moiré Brillouin zone (mBZ) in zero magnetic flux. But instead of dispersionless Landau levels, we recover the set of dispersive bands, with its band structure depending crucially on the twist angle (Fig.3).

\begin{figure}[t]
\includegraphics[width=0.9  \columnwidth]{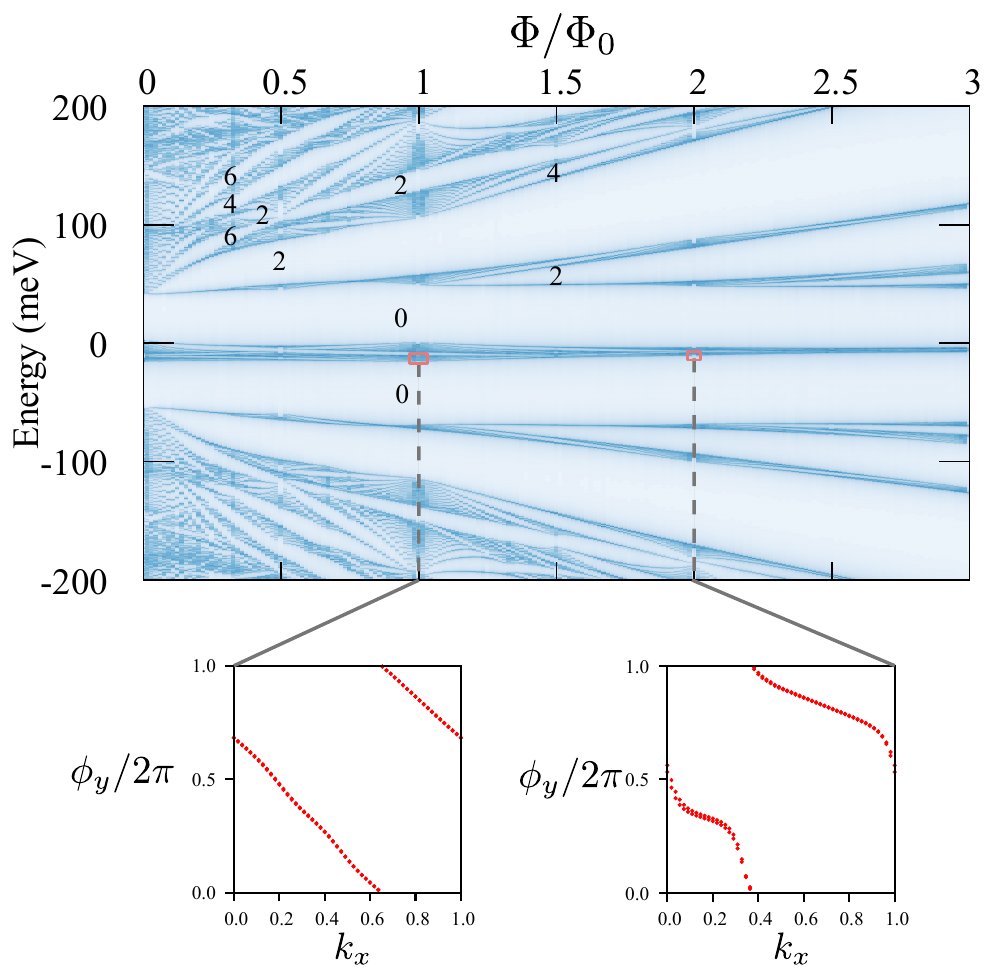}
\caption{Hofstader spectrum of magic-angle bilayer graphene { in magnetic flux (in units $h/e$). 
The digits in gaps indicate the in-gap Chern numbers, calculated through the edge modes counting. Additionally, the insets below  shows Wannier charge center (WCC) winding  at fluxes $h/e$ and $2 h/e$, calculated at the half filling of the flat bands (2 of 4 subbands are occupied). The WCC winding is nontrivial, revealing $|C|$$=$$2$ at half filling in the integer flux. 
}
}
\end{figure}

\textbf{Re-entrant magic angle spectra at integer magnetic flux.}  We start from considering the electronic band structure at an integer magnetic flux $\Phi = N h/e$. Such a flux provides the $2 \pi N$ circulation of magnetic vector potential, and hence reconstructs the moire Brillouin zone (mBZ);  magnetic translation operators commute at integer flux $[\hat T_{\vec a_1}, \hat T_{\vec a_2}] = 0$. This allows us to re-introduce momentum as a good quantum number and compute the electronic band structure starting from the tight binding model for magic-angle twisted bilayer graphene with atomic relaxations, modified with Peierls substitution, 
\begin{align}
t_{ij} \to  t_{ij} e^{- i\frac{e}{h} \int_{\vec r_j}^{\vec r_i} \vec A (\vec r')  d \vec r'} . 
\end{align}
  Comparing to the widely-used continuum models,\cite{Santos2007,BM2011,TKV} the accurate tight binding model  has an important advantage of addressing the magic-angle physics under realistic conditions of atomic lattice relaxations, proved to be indispensable in the experiments due to domain formation.\cite{Carr2018} Worth noting, the finite magnetic flux shifts the effective Brillouin zone (see Fig. 1), thus re-defining the positions of high-symmetry points. Otherwise, the original moire BZ and reconstructed magnetic moire Brillouin zone (mmBZ) have the same orientation and periodicity, which is the main technical condition to observe the magic angle phenomena.

We report that the characteristic flat band, the hallmark of the magic-angle graphene,  re-appears in \textit{every} integer magnetic flux, $\pm \Phi_0$, $\pm 2 \Phi_0$, $\pm 3 \Phi_0$... 
Figure 1 provides the electronic band spectrum at $\Phi = 0, \Phi_0, 2 \Phi_0$, while the band structures (and other properties) in stronger magnetic fluxes are given in the SM.\cite{SI} The first observation is that the flat band reappears exactly at the magic angle, while the higher bands are  dispersive (see Fig.3). We argue below that the magic-angle flat band (MAFB) in the integer  flux \textit{are not} a consequence of Landau level (LL) flattening since: \textbf{(i)} It has $|C|$=$2$ at half-filling; \textbf{(ii)} It demonstrates quantum geometry incompatible with  LL physics (Fig.5); and \textbf{(iii)} It becomes dispersive outside the magic angle. Worth noting, the strong magnetic fields restore the asymptotic particle-hole symmetry of the low energy states, which was moderately broken in the zero flux.

\begin{figure}[b]
\includegraphics[width=1 \columnwidth]{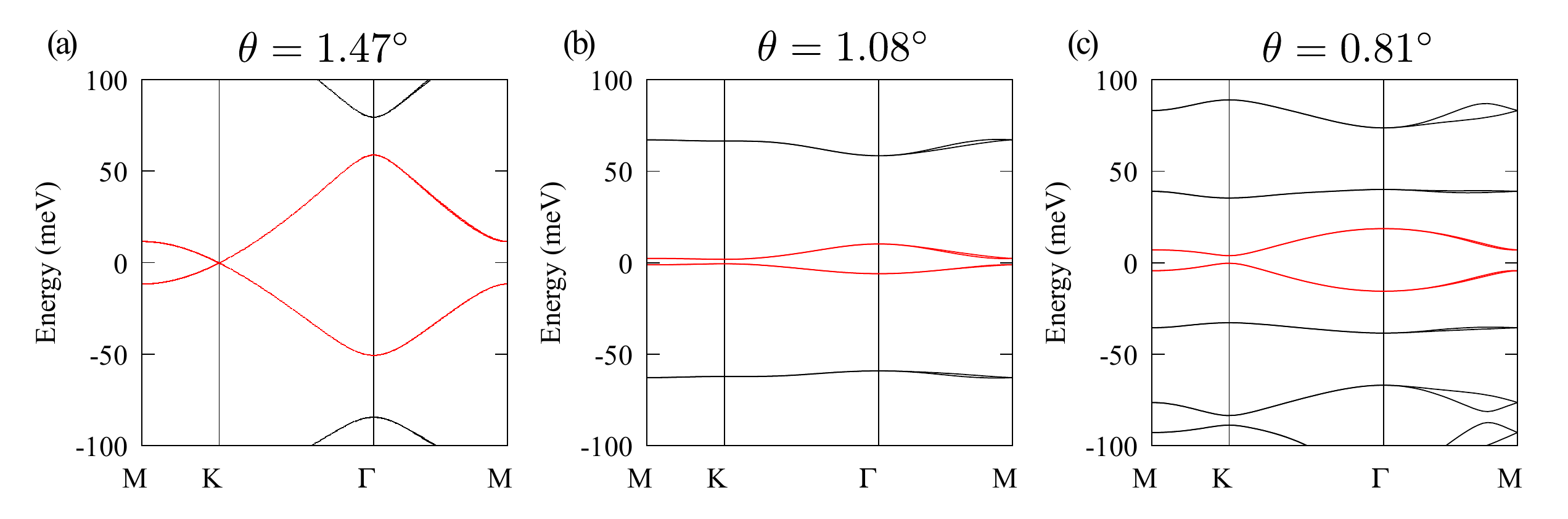}
\caption{
		Magic angle signature in integer magnetic flux $\Phi=h/e$ at different twists.
(a) Above the magic angle, $\theta = 1.47^{\circ}$. (b) At the magic angle, $\theta = 1.08^{\circ}$. (c) Below the magic angle, $\theta = 0.81^{\circ}$. We observe the similar behavior as the zero-flux TBG tuned outside of the magic angle.\cite{Bistritzer2011,TKV} }
\end{figure}

\textbf{Magnetic spectrum distinct from Landau levels}. We further address the properties of the electronic band spectrum versus magnetic field. For this, the characteristic quantity to calculate is the Hofstadter diagram, which traces the energy of electronic states allowed in quantized magnetic field, as a function of magnetic flux through the unit cell (Fig. 2). 
We observe that the flat bands at the integer flux are not stemming from Landau level Hofstadter physics, but rather from the magic angle physics of the TBG. To show that they are fundamentally different from the LLs, we calculate the Chern number through Wilson loop computation in the form of Wannier charge center winding, see Fig. 2.  
We find that the flat bands at integer $\Phi$ have  Chern numbers $|C|=2$ incompatible with Chern numbers of  Landau levels on the lattice ($|C|=1$). 
Furthermore, we fix magnetic flux to $h/e$, and investigate the change in electronic band spectrum versus the \textit{change in  twist angle} (see Fig. 3; Cases of higher integer flux are provided in SM.\cite{SI}).   While the spectrum is strongly dispersive outside of the magic angle (the bandwidth is approximately 100 meV at $1.47^{\circ}$ twist, at the magic angle $1.08^{\circ}$ the bandwidth is just $15$ meV, \textit{comparable} 
to the magic angle bandwidth in zero flux  (see also Fig. 1). We conclude that the magic angle physics is in its essence restored.

\begin{figure}[t]
\includegraphics[width=0.65  \columnwidth]{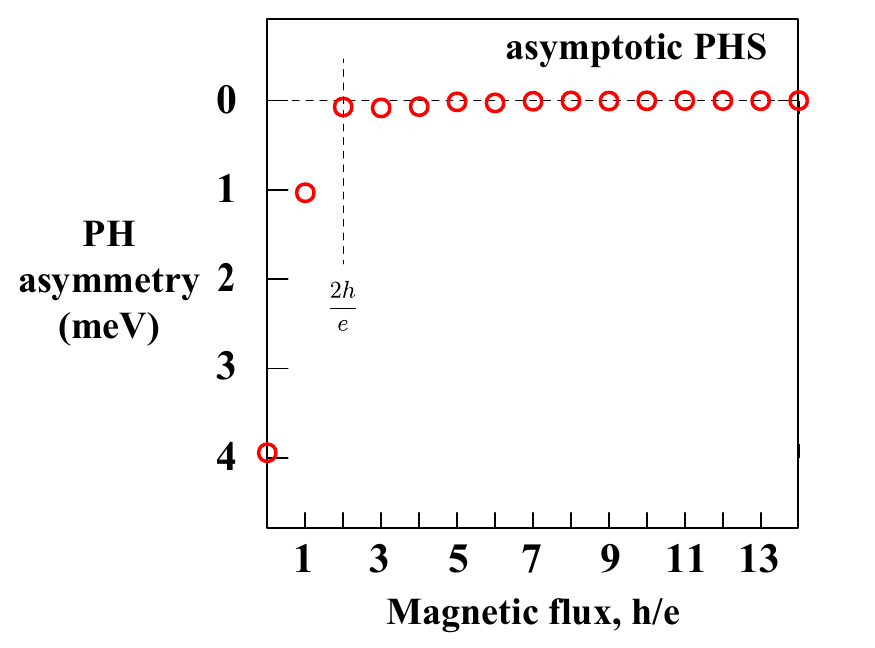}
\caption{
		Asymptotic particle-hole symmetry (PHS)  in strong magnetic flux
	 is emergent for $\Phi$$\ge$$2 h/e$. The low-energy PH asymmetry of the flat bands is determined as a maximum energy difference between the conductance and valence bands at neutrality, in meV. The PH asymmetry is maximal in zero flux ($\Phi=0$) resulting in nearly 4 meV. The PHS violation is however strongly suppressed with applied magnetic field, resulting to  1 meV at $\Phi=h/e$, and just 0.07 meV at $\Phi=2h/e$. The suppression of PH asymmetry roughly follows $\sim\exp[-\Phi/\Phi_*]$, with $\Phi_* \approx 0.7  h/e$.  }
\end{figure}

\textbf{Emergent particle-hole symmetry.}  We now focus on the low-energy numerical analysis of the flat bands. 
The main observations under strong magnetic fields is that  the low-energy electronic spectrum, featuring flat bands, acquires asymptotic particle-hole symmetry (Fig. 4) 
To quantify the particle-hole asymmetry (PHA), we track the energy difference maximum difference in the flat band energies between the occupied and free bands at neutrality, and investigate it versus magnetic field (Fig. 4). 
We observe that while PHS is generically  violated in zero flux (PHA = 4 meV), in strong magnetic fields this symmetry reemerges in its asymptotic form above $\Phi \ge 2 h/e$ (Fig. 4). 
Approximately, the suppression of PH asymmetry is observed as $\sim  \exp[-\Phi/\Phi_*]$, with $\Phi_* \approx 0.7  h/e$ (corresponding to $\approx 20$ T). 
This could be qualitatively understood in terms local tight binding hoppings, which due to strong magnetic fields oscillate rapidly in space; the system thus performs self-averaging which re-defines effective hopping parameters.  
In realistic TBG systems at $B=0$, with sublattice hoppings and lattice relaxations,  the Chern number in zero flux vanishes since the particle-hole symmetry (PHS) is violated on the atomistic level,  chiral symmetry (CS) is broken explicitly, and time reversal (TRS) is present.  
At integer fluxes, we have asymptotic PHS, broken TRS, broken CS, and while strictly speaking the system belongs to class $A$ topological insulator, its dynamics 
 flows towards class $C$, characterized by $2 Z$ (even-valued) topological invariants in 2D systems.\cite{Altland1997} This gives a plausible explanation of promotion of the $|C|$=$2$ Chern numbers in the flat bands at integer flux $|\Phi| \ge \Phi_0$, {once the two of four subbands are slightly gapped out (Fig. 1)}. 
We observe that the magic-angle TBG at zero flux  $\Phi =0$ and at integer fluxes $\Phi =\pm N \Phi_0$ belong to different topological classes, which should be taken into account for understanding recent TBG experiments.

\textbf{Non-trivial quantum geometric properties}.  We report that the magic-angle flat band in integer magnetic flux has non-trivial quantum geometry (Fig. 4),  distinct from Landau levels. The quantum geometry is defined for the Bloch states in the projective Hilbert space; it can be separated into real (diagonal) part which is Fubini-Study metrics, and imaginary (off-diagonal) part, with its components being Berry  curvature.\cite{Provost1980}  Numerically, we compute quantum-geometric tensor for flat bands in TBG by using its spectral representation\cite{Kruchkov2021b} 
\begin{align}
\mathfrak G_{ij} (\vec k) = \sum_{n,m}  \frac{
\langle u_{n \vec k}  |  \frac{\partial \mathcal H_{\vec k} }{\partial k_i}     |   u_{m \vec k}    \rangle_0  
\langle u_{m \vec k}  |  \frac{\partial \mathcal H_{\vec k} }{\partial k_i}     |   u_{n \vec k}    \rangle_0  
 } { (\varepsilon_{n \vec k} -  \varepsilon_{m \vec k})^2 }. 
\end{align}
We further introduce $\mathcal G_{ij}$=$\text{Re} \mathfrak G_{ij}$, $\mathcal F_{ij}$=$- 2 \, \text{Im} \mathfrak G_{ij}$. One can find a basis in which $\mathcal G_{ij}$ is diagonal and $\mathcal F_{ij} $ is off-diagonal. 
The plots for $\mathcal G_{xx}, \mathcal G_{yy}$ (Fubini-Study metrics)  {calculated at half-filling} are presented in Fig. 5. We observe that the re-entrant flat band has nontrivial quantum geometry within the mmBZ, as manifested in Figs. 5(a,b,c), which is not compatible with a Landau level quantum geometry {(the LL quantum geometry is constant in the whole Brillouin zone)}. 

For comparison, we plot the Berry curvature $\mathcal F_{xy}$ together with trace of Fubini-Study's  $\mathcal G_{ij}$ in Fig. 3d. 
We observe that the flat band closely follows the quantum-geometric condition for ideal flat bands\cite{Kruchkov2021a} $\text{Tr} \mathcal G_{ij}$=$\mathcal F_{xy} $. 
It is certainly interesting that this condition is satisfied almost exactly in the regions of mmBZ, where the band flatness is pronounced in terms of vanishing Fermi velocity (around the K points of mmBZ).  The deviation to this quantum-geometric bound $
\text{Tr} \mathcal G_{ij}$=$\mathcal F_{xy} 
$ are observed in the regions of mmBZ with finite dispersion and significant $v_k=\partial{\varepsilon_k}/\partial{k}$ caused by broken CS of the tight binding calculations.  The criterion  $
\text{Tr} \mathcal G_{ij} =  F_{xy}, 
$ 
 tests the closeness of a realistic flat band in TBG to flat band idealization through holomorphic/meromorphic representation of the flat band wave functions.\cite{Kruchkov2021a} 
 However, since total $|C|=2$, the relevant toy model for TBG in integer flux cannot be represented by solely a  holomorphic representation of the quasi-LLL TBG (found in Ref.\cite{TKV}); one needs to 
take meromorphic flat band contributions into account.\cite{Popov2021}

We  calculate the distribution of the Berry curvature $F_{xy}$ within the mBZ (the cut is shown in Fig. 3d, the density plots are in SM), which reveals non-homogeneous structure, which is  not consistent with the homogeneous Berry curvature of the generic Landau levels (LLs are "Berry flat").  This yet again provides arguments in the support of the magic-angle physics re-entrance in integer magnetic flux. Finally, we check numerically that the Berry flux $F_{xy}$ encompassed by mmBZ sums up to $|C|$=$2.0 \pm 0.07$ {at the half filling (see Figs. 1,2)}.

\textbf{Quantum-geometric transport.} As was first introduced by Peotta and T\"{o}rm\"{a}\cite{Peotta2015}, the nontrivial quantum-geometric tensor $\mathfrak G_{ij}$ leads to the the finite superfluid current $J_{i}$$=$$- D_{ij} A_j$ even in the limit of perfectly flat band (here $D_{ij} $ is the superfluid weight). This argument is now understood to apply directly to TBG in zero flux, where the $\text{Tr} \mathcal G_{ij}$ is nonzero {due to hidden nontrivial topology of the flat bands}. It was reported with different methods\cite{Hu2019,Xie2020,Julku2020} that the quantum geometric tensor (QGT) contribution to the superfluid weight $D_S$ in TBG is if not dominant, than at least commensurate with the conventional contributions, thus leading to the BTK transition temperature estimate $T_{\text{BKT}} \sim \sum_{\vec k} \text{Tr} \mathcal G_{ij}$.   The QGT contribution holds for different symmetries of the order parameter, and the argument is valid beyond the mean field.\cite{Wang2020qg} The essential physics is captured by 
Bogoliubov-de-Gennes Hamiltonian
\begin{align}
\mathcal H_{\text{BdG}}=
\left(
\begin{array}{cc}
H_{\vec k} & \Delta_{\vec k} \\
\Delta^*_{-\vec k} & H^*_{-\vec k}
\end{array}
\right)
\end{align}
Without loss of generality, we consider superconducting order $\Delta_{\vec k} = \Delta$. 
The superfluid weight is then calculated within Kubo formalism through the current-current correlators. 
 We explicitly calculate $\mathcal G_{ij}$ numerically (Fig. 4) with consequent mmBZ integration at half filling (for two of four flat bands occupied, we have $\text{Tr} \mathcal G_{ij} \approx {2.8}$), to obtain at the  {superfluid weight maximum positioned at the middle of the composite flat band with $|C|=2$,}
\begin{align}
D_{xx} = \frac{2 e^2}{\hbar^2} \Delta \sum_{\vec k} \text{Tr} \mathcal G_{ij} (\vec k) \approx (5.6 \pm  0.1) \frac{e^2}{\hbar^2} \Delta .
\end{align}
Symmetry $D_{xx}$$=$$D_{yy}$ is assumed. {Here we took into account factor $\sqrt{\nu (1 - \nu)}$  (in notations of Ref.\cite{Peotta2015}), where $\nu$ is the filling factor of the composite flat band (indexed by  $C=-2$ in Fig. 1)}.  

We can further make estimates for the BKT transition temperature,\cite{Berezinskii1971,Kosterlitz1973,Nelson1977} indicating the disappearing of the phase coherence of superconducting order from expression  $\pi \hbar^2 D (T_*) /8 e^2 T_*=1$. The order-of-magnitude estimate gives $T_* \sim {\hbar^2 D (0)}/{e^2} \sim \Delta$. The remaining question is of course, what is the value of $\Delta$, which should be found self-consistently by solving Gorkov equations in magnetic field, or through indirect experimental data---and it is beyond the scope of this paper. For a rough estimate, even $\Delta \sim 0.1$ meV will  give a physically relevant $T_* \sim 1$ K. Whether or not the superfluid order is re-entrant in strong magnetic flux remains open,\cite{Chaudhary2021,Shaffer2021} however it would not be surprising in the view of the reported Pauli-limit violation and re-entrant superconductivity in strong magnetic parallel fields in twisted graphene multilayers.\cite{Cao2021,Park2021b,Zhang2021} Note the magnetic field may or may not change the symmetry of the superconducting order (resulting into different $\Delta_{\vec k}$), however the quantum-geometric superfluid weight will still remain finite due to the nontrivial geometry of the underlying flat bands.

\begin{figure}[t]
\includegraphics[width=1  \columnwidth]{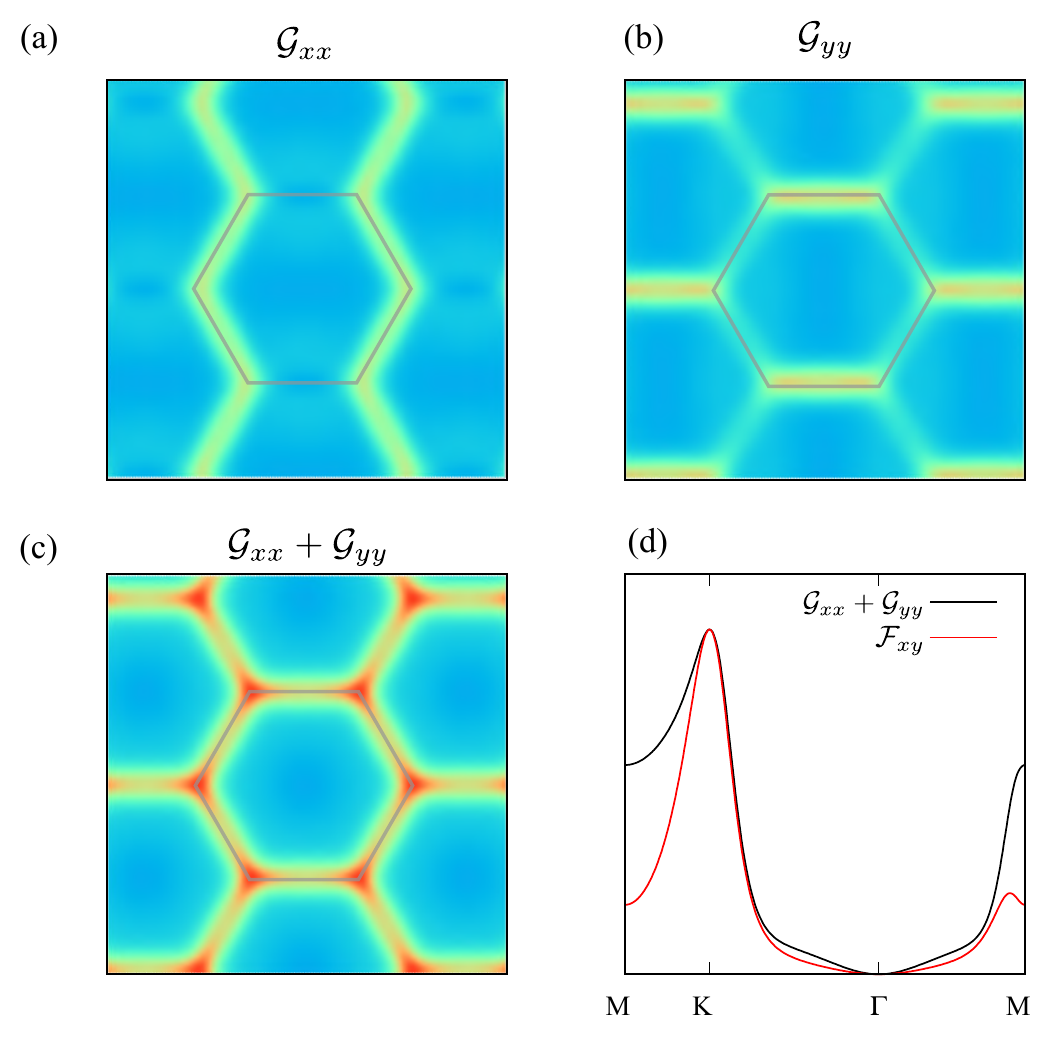}
\caption{Nontrivial quantum metrics in integer magnetic flux $\Phi=h/e$, calculated at half-filling (two of four flat bands are occupied). (a-c) components of Fubini-Study metrics; (c) shows that $\text{Tr} \mathcal G_{ij}$ is gauge-invariant (d) Comparison between trace of Fubini-Study tensor $\text{Tr} \mathcal G_{ij}$ and Berry curvature $\mathcal F_{xy}$, which probes how close the band flatness is  to the perfect flatness through Eq.\eqref{ideal}.  \vspace{-2 mm}}
\end{figure}

\textbf{Conclusions}.
We conclude that there is a re-entrant magic angle physics in twisted bilayer graphene 
at	every integer magnetic flux quanta
 $\pm \Phi_0$, $\pm 2 \Phi_0$, $\pm 3 \Phi_0$, \textit{etc.}, through the moiré cell. 
To date, the practical importance represents the first magnetic quantum $\pm \Phi_0$, which at twist angle $1.08^{\circ}$ corresponds to experimentally-achievable fields of 28 Tesla. We confirm with accurate atomistic calculations, incorporating lattice relaxation effects, that at such fields the magic-angle phenomena re-emerge. This, in particular, could be seen through the re-emergence of very flat bands at the magic angle  distinct from Landau levels, while beyond the magic angle this physics breaks down, similar to the zero-flux case.\cite{Bistritzer2011,TKV} These flat bands {at half filling} carry nontrivial  Chern numbers ($|C|$=$2$)  and nontrivial quantum geometry (Fubini-Study metrics),  and are fundamentally different from conventional Landau levels.  We conjecture that, similar to TBG in the zero flux, there is a nonvanishing contribution to the superfluid weight coming from the re-entrant quantum geometric properties, and in the flat topological bands this contribution is significant. Due to the strong quantum geometry of the TBG flat bands in integer flux ($\sum_{\text{BZ}}\text{Tr} \mathcal G_{ij} \approx 3$), and the estimated BKT temperature is in order of the gap $T_* \approx \Delta$, which gives values significantly elevated with regard to the conventional superconductivity in geometrically-trivial dispersive bands.  
The behavior of superfluid order parameter beyond the conventional Pauli limit (towards integer magnetic flux) is a subject for further research.

\

\textbf{Acknowledgments.} The authors thank Ady Stern, Yarden Sheffer, Luiz Santos, Gaurav Chaudhary,  and Marta Brzezińska for useful discussion. {The authors thank B. Andrei Bernevig, Jonah Herzog-Arbeitman, Aaron Chew for further discussion.}
The project was supported by the Branco
Weiss Society in Science, ETH Zurich, through the research grant on flat bands, strong interactions and SYK
physics, and by the Swiss National Science Foundation,
grants No. 172543.  Computations
have been performed at the Swiss National Supercomputing Centre (CSCS) under project s1008 and the facilities
of Scientific IT and Application Support Center of EPFL.

\bibliography{Refs}

\end{document}